# A highly efficient method for second and third harmonic generation from magnetic metamaterials

**Iman Sajedian[1,2], Inki Kim[2], Abdolnasser Zakery[1] and Junsuk Rho[2,3,*]**

[1]*Department of Physics, College of Science, Shiraz University, Shiraz 71454, Iran*

[2]*Department of Mechanical Engineering, Pohang University of Science and Technology (POSTECH), Pohang 37673, South Korea*

[3]*Department of Chemical Engineering, Pohang University of Science and Technology (POSTECH), Pohang 37673, South Korea*

[*] *jsrho@postech.ac.kr*

Second and third harmonic signals have been usually generated by using nonlinear crystals, but that method suffers from the low efficiency in small thicknesses. Metamaterials can be used to generate harmonic signals in small thicknesses. Here, we introduce a new method for amplifying second and third harmonic generation from magnetic metamaterials. We show that by using a grating structure under the metamaterial, the grating and the metamaterial form a resonator, and amplify the resonant behavior of the metamaterial. Therefore, we can generate second and third harmonic signals with high efficiency from this metamaterial-based nonlinear media.

## INTRODUCTION

Metamaterials have achieved many breakthroughs in today's optics by their special properties, such as negative refractive index[1-4], invisibility cloaking[5-8], super-resolution imaging[9-12], metasurfaces[13-16]. Many nonlinear effects have also been shown and proposed based on metamaterials. Like using nonlinear photonic elements such as diodes for generating nonlinear effects [17,18]. Second and third harmonic generation[19-21], solitons[22], and tunable nonlinear metamaterials by using liquid crystals[23].

In this paper, we introduce a highly efficient method for generating second and third harmonic signals from metamaterials. The current loop in the magnetic resonance of metamaterials generates second and third harmonic signals in two different polarizations[?]. Here we show how using a grating under the metamaterial amplifies these currents, which leads to the amplification of the second and the third harmonic generation from metamaterials

## MATERIALS AND METHODS

Magnetic metamaterials show their nonlinear behavior in their magnetic resonance. In order to find the resonance regions, we should obtain the metamaterial absorption spectra (see Figure 1(a)). We illustrate our idea of highly efficient nonlinear signal generation, by focusing on the magnetic resonance of nanostrip metamaterials[24]. The maximum in the absorption spectra gives us the magnetic resonance frequency (see Figure 1(a)). In the magnetic resonance frequency, the normal component of the magnetic field induces a magnetic dipole inside the metamaterial, which induces a polarization current loop in the metallic parts (see Figure 1(b)).

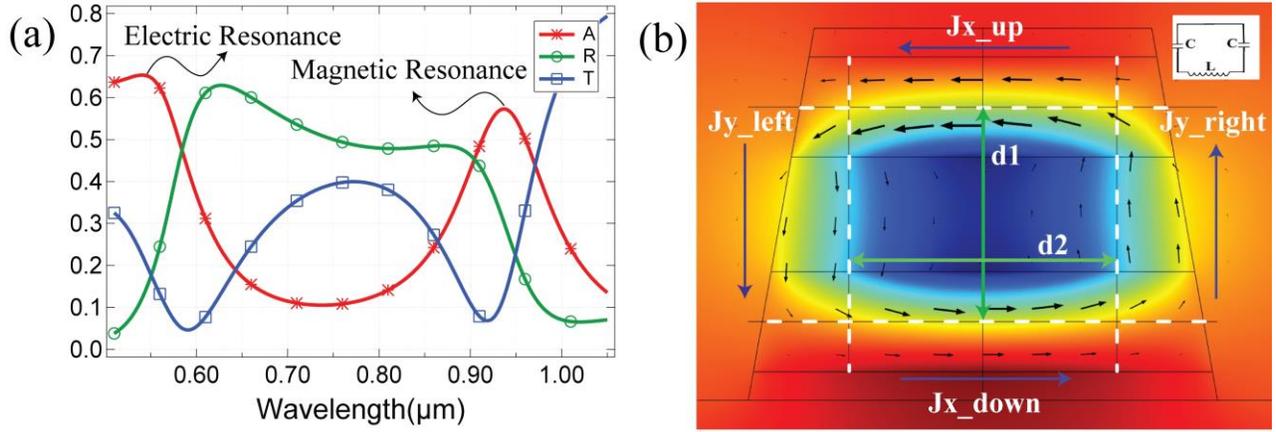

Figure 1. (a) Transmission, reflection and absorption spectra of nanostrip. The peaks in the absorption curve determine the resonant frequencies. (b) Magnetic field distribution and electric field displacement (black arrows) in magnetic resonant frequency. The external magnetic field induces a current loop which can be divided into two orthogonal groups. The current loop components are measured on the white dashed lines. The inset shows the equivalent circuit.

To see the effect of currents in generating nonlinear signals in metals, we investigate the linear and nonlinear behavior of metals. The linear behavior of metals can be obtained by using the Drude model, which is the equation of motion of free electrons in metals[25]:

$$m\frac{\partial^2 \vec{r}(t)}{\partial t^2} + m\Gamma \frac{\partial \vec{r}(t)}{\partial t} = -e\vec{E_0}e^{-i\omega t} \quad (1)$$

In the above formula, the right-hand side term is the driving force caused by the exciting wave, and the middle term is the damping force. In order to get the nonlinear behavior, we should add two physical effects, which are electric and magnetic components of Lorentz force and convective derivative of the electron velocity field[26]:

$$\frac{\partial \vec{J_p}}{\partial t} = -\Gamma \vec{J_p} + \varepsilon_0 \omega_p^2 \vec{E} + \sum_k \frac{\partial}{\partial r_k}\left(\frac{\vec{J_p}J_{pk}}{\omega_p^2 m\varepsilon_0/e - \rho}\right) - \frac{e}{m_e}\left[\rho\vec{E} + \vec{J_p} \times \vec{B}\right] \quad (2)$$

where $\omega_p = \sqrt{\frac{n_0 e^2}{\varepsilon_0 m}}$ is the plasma frequency, $\vec{J_p} = \frac{\partial \vec{P}}{\partial t}$ and $\vec{p(t)} = ne\vec{r}(t)$ are the polarization current and the polarization field. So by amplifying the polarization currents we can see nonlinear behavior from metals. In the magnetic resonance frequency of metamaterials, the polarization current $\vec{J_p}$ reaches its maximum, and if we increase the initial power high enough we can see the nonlinear effects.

Now we investigate this specific metamaterial more closely. Nanostrips have inversion symmetry in the x-direction (parallel to the gap), and no inversion symmetry in the y-direction (perpendicular to the gap). So the x component of polarized current $J_x$ generates third harmonic signal, and the y component of polarized current $J_y$ generates second harmonic signal.

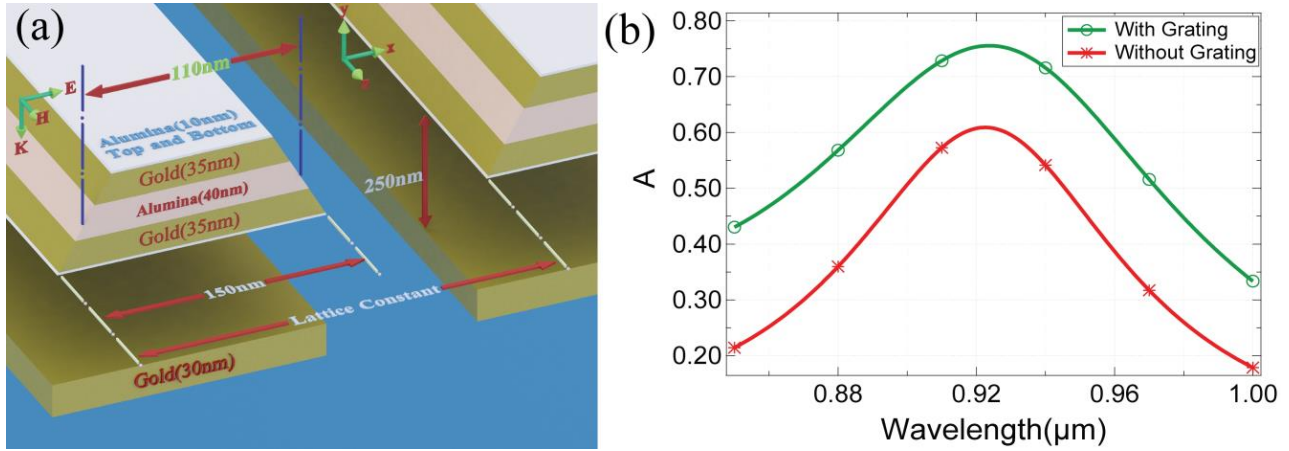

Figure 2. (a) Geometrical parameters of the model. The model is a 5-layer trapezoid placed on a glass substrate upon a grating, which is assumed to be very long. The applied field has TM polarization, which means the magnetic field is normal to the metamaterial's cross section, and the electric field is parallel to the gap. The grating has the same width as the metamaterial and is embedded in the glass substrate. The optimum separation between the grating and the nanostrips and the optimum height of the grating is determined by simulation. (b) Absorption spectra of nanostrips with and without grating. In the metamaterial with the grating case, we have a higher absorption in the magnetic resonance frequency.

By using a grating under the metamaterial, the metamaterial and the grating form a Fabry-Perot like cavity, and so the field inside the metamaterial is amplified. This means that the polarization currents are amplified too and so we have more efficient nonlinear effects. The reason that we used grating instead of a grounded metallic plane (which would act as a mirror) is that the nonlinear effects are more powerful in the transmitted wave than in the reflected wave, so by using the grating we have both the resonating effect and the transmission.

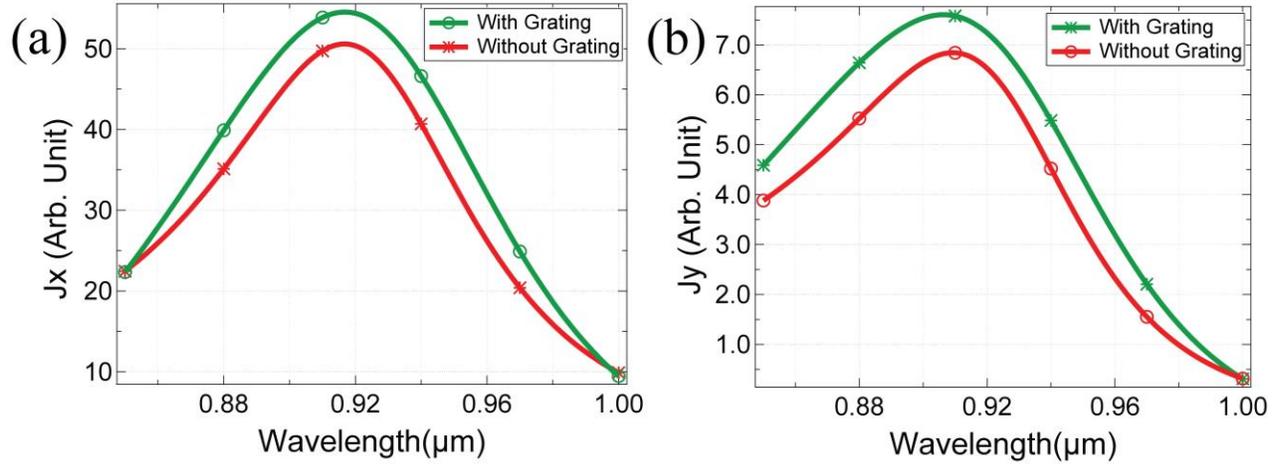

Figure 3. (a) polarization current, x-component (Jx_tot) and (b) polarization current, y-component (Jy_tot) in the magnetic resonance of the nanostrips with and without using the grating. Both of the $J_x$ and $J_y$ peaks are amplified, which means we should see higher amplitudes of nonlinear generation from the metamaterial in these frequencies.

**RESULTS AND DISCUSSION**

The model that we used, consists of a five layers model for the nanostrip and a grating. Two 10nm thick alumina ($Al_2O_3$) on the top and the bottom of the nanostrip and a gold- alumina -gold in the middle of the structure (see Figure 2(b)). The thickness of the gold layers is 35nm and the thickness of the middle alumina layer is 40nm. The space between the grating and the nanostrips is 250 nm and the height of the grating is 30nm. The optimum values are obtained by simulation. For the simulations, we used COMSOL Multiphysics. As for the simulation parameters we used continuity boundary conditions on the left and right boundaries and non-reflecting conditions in the top and bottom boundaries. We excited the model with a gaussian wave with TM polarization from the top boundary. The optical parameters used are n=1.5 for glass, n=2.25 for alumina[27], n=2.5074, $\omega_p$ = 13.67×10$^{15}$ 1/s, $\Gamma$=0.0648×10$^{15}$ 1/s, $\varepsilon_\infty$=9.84 for gold[25,28]. For measuring SH and TH signals, we used a probe at the end of the simulation region, which means we measured the transmitted wave.

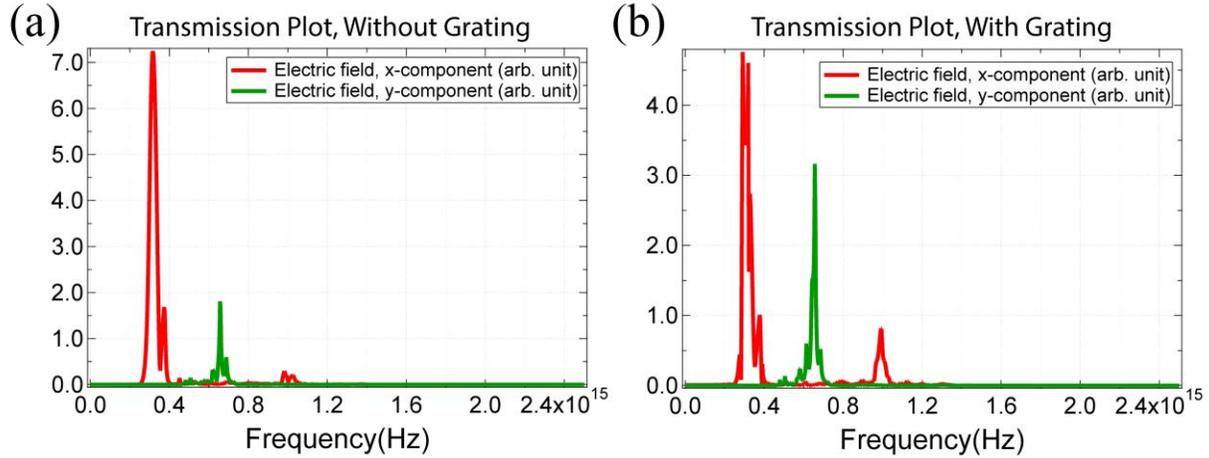

Figure 4. (a) Harmonic generation in the transmitted wave without using the grating, in the magnetic resonance of the metamaterial. (b) Harmonic generation in the transmitted wave with using the grating, in the magnetic resonance of metamaterial. After using the grating, the amplitude of the second harmonic signal is more than doubled and the amplitude of the third harmonic signal is approximately quadrupled.

To show the effect of the grating on the metamaterial, first we measure the absorption in the magnetic resonance in both cases. As can be seen from Figure 2(b) absorption is much higher while using the grating. Next, we measure the average currents on four lines on the upper ($J_{x,up}$) and the lower ($J_{x,down}$) metallic parts, and the left ($J_{y,left}$) and the right ($J_{y,right}$) metallic parts of the nanostrips(see Figure 1(b)). Measuring these currents shows the effect of grating on each of the second and third harmonic generation. Considering the phase difference, we have the following quantities[?]:

$$J_{x,tot} = J_{x,up} \times \exp(-ik_0 d_1) + J_{x,down} \quad (3)$$

$$J_{y,tot} = J_{y,left} \times \exp(-ik_0 d_2) + J_{y,right} \quad (4)$$

As can be seen in Figure 3(a,b) after using the grating both of $J_{x,total}$ and $J_{y,total}$ quantities have been amplified. This is in consistency with the higher absorption after using the grating (see figure 2(b)). Measuring the above quantities shows the effect of grating on harmonic generation (see Figure 3(a,b)). As can be seen in Figure 4(a,b), the amplitudes of second harmonic signals in the case of the metamaterial with grating, is almost doubled and the amplitudes of third harmonic signals is almost quadrupled in comparison to the case of no grating.

**CONCLUSIONS**

In summary, we proposed a new method to enhance the efficiency of harmonic generation in metamaterials in their magnetic resonance frequency. We showed how adding a grating to the metamaterial, forms a resonating cavity that amplifies the polarization currents in the metallic parts of the metamaterial. By investigating this idea in the nanostrips metamaterials we saw a doubled amplitude in SHG and a quadrupled amplitude in THG in comparison to the no grating case.